\documentclass[submission,copyright]{eptcs}
\providecommand{\event}{WIVACE 2013} 
\usepackage{breakurl}             
\usepackage{graphicx}

\title{Signed Networks, Triadic Interactions and the Evolution of Cooperation}
\author{Simone Righi \qquad\qquad K\'{a}roly Tak\'{a}cs
\institute{MTA TK "Lend\"{u}let" Research Center for Educational\\ and Network Studies (RECENS).\\ Budapest (Hungary)}
\email{simone.righi@tk.mta.hu  \quad\qquad karoly.takacs@tk.mta.hu}
}
\def\titlerunning{Signed Networks, Triadic Interactions and the Evolution of Cooperation}
\def\authorrunning{S. Righi, K. Tak\'{a}cs}

\begin{document}
\maketitle

\begin{abstract}
We outline a model to study the evolution of cooperation in a population of agents playing the prisoner's dilemma in signed networks. We highlight that if only dyadic interactions are taken into account, cooperation never evolves. However, when triadic considerations are introduced, a window of opportunity for emergence of cooperation as a stable behaviour emerges.
\end{abstract}

\paragraph{Introduction.}In this paper we examine how cooperation can be supported by negative ties and triadic interactions. Previous research has demonstrated that cooperation is more likely to evolve in social dilemma games, such as the PrisonerÕs Dilemma, if the game is played in networks \cite{Hauertetal2004, Ohtsuki2006, Santos2006, Wang2008}. Besides sparseness \cite{Nowak2005, Ohtsuki2006, Santos2006}, the structure of the network is also important because on top of direct ties, indirect relations also control behavior and contribute to the establishment of cooperation through reputation mechanisms, such as image scoring \cite{Wedekind2000}. These ideas have been fruitful in explaining why and under which conditions can cooperation evolve and be stable among humans. In these studies however, only positive relations between players are analyzed.

Negative ties could potentially be better enforcers of cooperation than positive ones. At the dyadic level, social psychological mechanisms of vengeance and anger are manifested through the use of powerful trigger strategies in the repeated PrisonerÕs Dilemma game \cite{Axelrod1984, Axelrod1997, AxelrodHamilton1981, Trivers1971}. Another mechanism is selectivity that prescribes cooperation with those who are liked and defection with those who are disliked. Experiments demonstrate that subjects play differently in social dilemma games with related and unrelated others (e.g., \cite{Ferraro2007}). There is also evidence of homophilous selection of partners that can stabilize cooperative regimes \cite{Yamagishi1996}.
Furthermore, we hypothesize that when interactions happen on a triadic basis, cooperation can increase. For instance, if A and B are friends, and C is their common enemy, the presence of the negative ties towards C could strengthen cooperation between A and B.
Signed relations could reliably signal cooperative or non-cooperative intentions, and hence could guarantee the spread of appropriate reputational information that result in confirming expectations about cooperation.

\paragraph{Model.}We incorporate both dyadic and triadic interaction mechanisms in an agent-based model in which the (random) signed network  and cooperation co-evolve. Our model is inspired by the tradition of computational evolutionary game theory models where the more successful strategies tend to diffuse in the population. We analyze the evolutionary success of different strategies that are defecting or cooperating and sensitive to the sign of links from their partners to others to a different way and extent. Indeed, each agent in our setup can be of one of the following three types:
\begin{itemize}
\item Unconditional Defector (UD): which will always defect.
\item Conditional Player(CO): whose strategy choice is influenced by the sign of his ties. He cooperates if he likes the opponent (positive tie) and he defects if he does not like the opponent (negative tie). 
\item Unconditional Cooperator (UC): It cooperates always regardless to the sign of his ties. 
\end{itemize}
Starting with a population composed of a similar number of agents for each type we iterate the following steps until an equilibrium is reached:
\begin{enumerate}
\item Two \textit{(or three in the case of triadic interactions)} connected agents are selected randomly. 
\item The prisoner's dilemma is played between the two \textit{(or three, playing the game between each couple of the triad)} agents selected and the total payoff is calculated for each of them\footnote{If a triad is selected all three agents need to have connections with each other so that everyone will play exactly twice.}.
\item The signs of the network ties involved in the interactions evolve as consequence of the actions of the agents in step 3. In particular:
\begin{itemize}
\item if both agents played the \textit{Defection} strategy then the sign of their tie is turned negative.
\item if both agents played the \textit{Cooperation} strategy then the sign between them is turned positive.
\item if one of the agent \textit{Cooperates} while the other \textit{Defects} then the tie is turned positive with probability $P_{pos}$ and negative with probability $P_{neg}$.
\end{itemize}
\item The agent with higher payoff has a probability $P_{inv}$  of ÒinvadingÓ the agent with lower payoff (in case of equality each agent has the same probability of invasion\footnote{If there is a payoffs order $a>b>c$, then only $a$ can invade one of the other agents. However, if the payoffs order is of the type $a=b>c$ than either $a$ or $b$ can be selected. The selected agent will then randomly choose to invade one of the remaining players of the triad randomly.}).
\end{enumerate} 
Studying this setup in the two different cases in which the Prisoner's dilemma is played in couples and in triads we obtain the preliminary results outlined in the next paragraph\footnote{For reasons of space, in this paper we limit our analysis to a conceptual analysis.}
 
\begin{figure}[h]
\centering
\includegraphics[width=7cm]{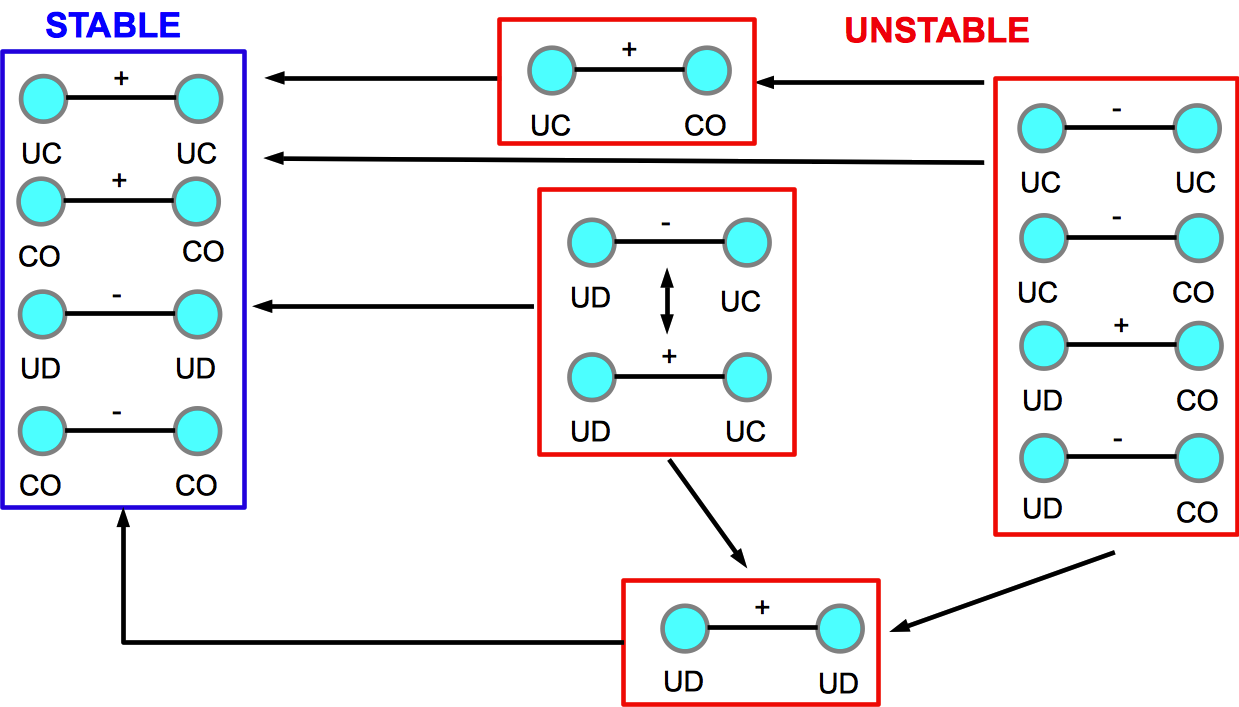} \\
\vspace{0.5cm}
\includegraphics[width=5.7cm]{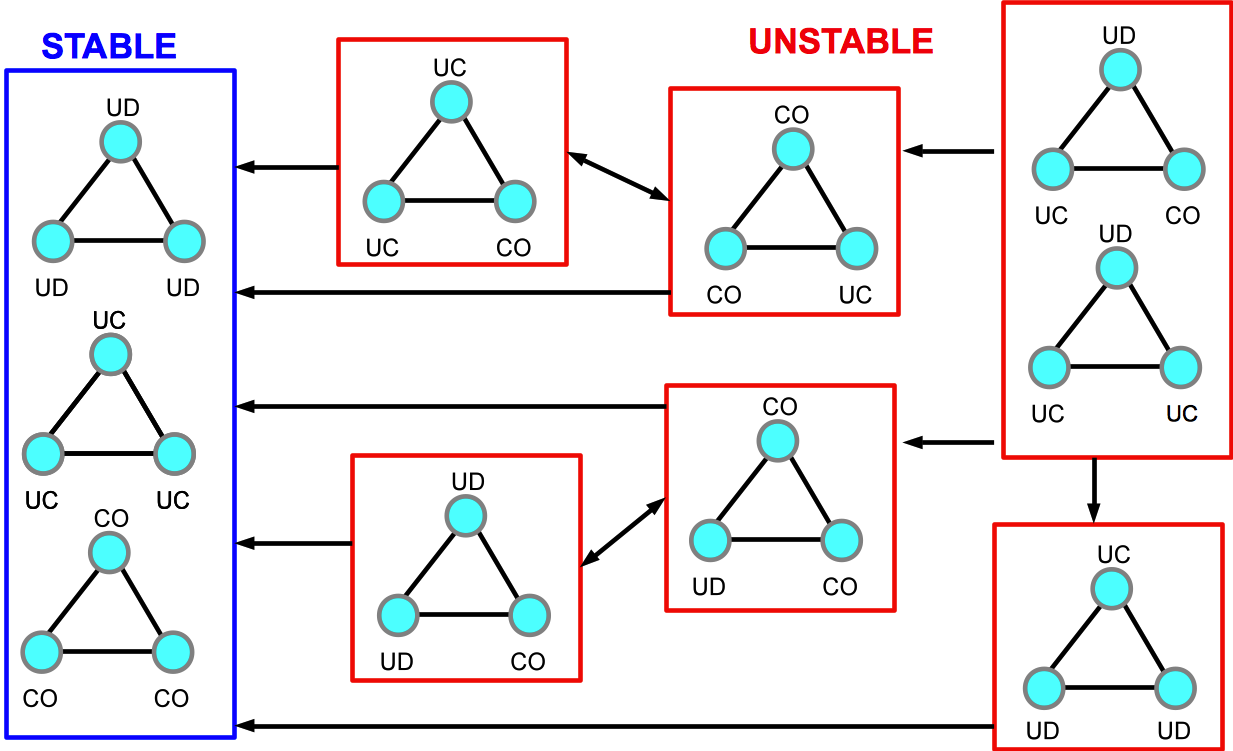} \quad \quad
\includegraphics[width=5.7cm]{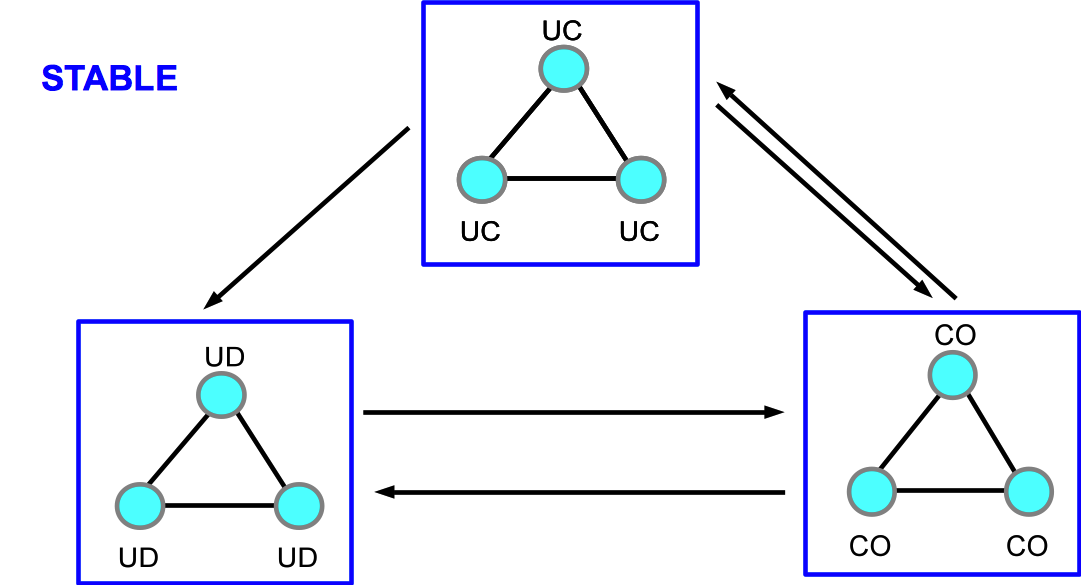} 
\caption{The \textbf{first panel} shows the effects of model's dynamics on each signed pair for the case in which only dyadic interactions are considered. The arrows between dyads indicate the possible transformations from one pair to another allowed by the model's dynamics. Only the four pairs circled in blue (the four left ones) are stable. The \textbf{lower left panel} repeat the same exercise for triadic interactions. The \textbf{lower right panel} shows that none of the triads which are stable in isolation is dominated by any other. The signs are omitted from the figure and each triad represent all possible triads with the combination of agents' types indicated, regardless of the signs that link them. The arrows between triads indicate possible transformations from one triad to another, accounting for the possibilities given by different network signs, allowed by the model's dynamics.}.   
 \label{fig:dyadicandtriadics}
 \end{figure}

\paragraph{Outline of the results.}In the presence of negative ties, cooperation cannot be sustained in a random network where only dyadic considerations affect the agents' behavior. The reason is described by the first panel of Figure \ref{fig:dyadicandtriadics}. Here we consider the stability, in isolation, of each possible pair of agents and sign of the tie that links them. If a pair is unstable then the dynamics of the model will transform it in some other pair. In presence of only dyadic interactions only the pairs circled in blue\footnote{Namely: UC+UC, CO+CO, UD-UD and CO-CO.} are stable. 
However, since the prisoner's dilemma is not played in isolation but in a population of agents, one must also consider the stability of the dyads to \textit{invasion} by other types of agents. Indeed, as UD always dominates UC in a dyadic comparison, the dyad UC+UC is only stable in isolation. Moreover, UD dominates CO in a dyadic comparison if they are linked by a positive tie, they are equivalent otherwise. So if $P_{neg}>0$ then $CO+CO$ and $CO-CO$ are also only stable in isolation. Generalizing this result to the whole network we can conclude that the only stable configuration for the case of dyadic interactions is a population of \textit{Defectors} linked by a network of negative signs.

Introducing triadic interactions enlarges the window of opportunity for the diffusion of cooperative behavior allowing for the emergence of stable cooperative clusters. Indeed, with the same strategy used above\footnote{Signs are not reported in the image so to simplify the drawing.}, we obtain the updating dynamics of the lower left panel of Figure \ref{fig:dyadicandtriadics}: all triads formed of agents of the same types are stable in isolation. Moreover, as can be seen on the right panel of Figure \ref{fig:dyadicandtriadics} none of this triads dominates any other one. This allows us to conclude that if interactions happen in triads it is possible to observe the emergence of some degree of cooperation in an evolutionary prisoner's dilemma played on signed networks. The actual emergence of cooperation depends on the values assumed by the transition probabilities and from the network topology assumed. We are currently designing simulations to corroborate the conceptual analysis provided here.
\nocite{*}
\bibliographystyle{eptcs}
\bibliography{RighiTakacsWivace2013}
\end{document}